\documentstyle[amssymb,epsfig,aps,pre]{revtex}

\tightenlines
\def\beq{\begin{equation}}
\def\beqa{\begin{eqnarray}}
\def\eeq{\end{equation}}
\def\eeqa{\end{eqnarray}}

\begin{document}
\title{Comment on ``Theory and computer simulation for the equation of state of
additive hard-disk fluid mixtures'' }
\author{M. L\'{o}pez de Haro\cite{mariano}}
\address{Centro de Investigaci\'on en 
Energ\'{\i}a, UNAM,\\
Temixco, Morelos 62580, M\'{e}xico}
\author{S. B. Yuste\cite{santos}}
\address{Institute for Nonlinear Science and Department of Chemistry, University 
of California,\\
San Diego, La Jolla, CA 92093-0340, U.S.A.}
\author{A. Santos\cite{andres}}
\address{Department of Physics, University of Florida,\\
Gainesville, FL 32611, U.S.A.}
\date{\today}
\maketitle

\begin{abstract}
A flaw in the comparison between two different theoretical equations of
state for a binary mixture of additive hard disks and Monte Carlo results,
as recently reported in C. Barrio and J. R. Solana, Phys. Rev. E {\bf 63},
011201 (2001), is pointed out. It is found that both proposals, which
require the equation of state of the single component system as input, lead
to comparable accuracy but the one advocated by us [A. Santos, S. B. Yuste,
and M. L\'{o}pez de Haro, Mol. Phys. {\bf 96}, 1 (1999)] is simpler and
complies with the exact limit in which the small disks are point particles.
\end{abstract}

In a recent paper, Barrio and Solana\cite{BS01} proposed an equation of
state (EOS)\ for a binary mixture of additive hard disks. Such an equation
reproduces the (known) exact second and third virial coefficients of the
mixture and may be expressed in terms of the EOS of a single component
system. They also performed Monte Carlo simulations and found that their
recipe was very accurate provided an accurate EOS for the single component
system (in their case it was the EOS proposed by Woodcock\cite{W76}) was
taken as input. The comparison with other EOS for the mixture available in
the literature indicated that their proposal does the best job with respect
to the Monte Carlo data. Among these other EOS for the binary mixture, only
the one introduced by us a few years ago\cite{SYH99} also shares with Barrio
and Solana's EOS the fact that it may be expessed in terms of the EOS for a
single component system. What we want to point out here is that the
comparison made in Ref. \cite{BS01} is flawed by the fact that it was not
performed by taking the {\em same} EOS for the single component system in
both proposals.

Let us consider a binary mixture of additive hard disks of diameters $\sigma
_{1}$ and $\sigma _{2}$. The total number density is $\rho $, the mole
fractions are $x_{1}$ and $x_{2}=1-x_{1}$, and the packing fraction is $\eta
=\frac{\pi }{4}\rho \langle \sigma ^{2}\rangle $, where $\langle \sigma
^{n}\rangle \equiv \sum_{i=1}^{2}x_{i}\sigma _{i}^{n}$. Let $Z=p/\rho k_{B}T$
denote the compressibility factor, $p$ being the pressure, $T$ the absolute
temperature, and $k_{B}$ the Boltzmann constant. Then, Barrio and Solana's
EOS for a binary mixture of hard disks, $Z_{\text{m}}^{\text{BS}}(\eta )$,
may be written in terms of a given EOS for a single component system, $Z_{%
\text{s}}(\eta )$, as 
\begin{equation}
Z_{\text{m}}^{\text{BS}}(\eta )=1+\frac{1}{2}(1+\beta \eta )\left( 1+\xi
\right) \left[ Z_{\text{s}}(\eta )-1\right] ,  \label{1}
\end{equation}
where $\xi \equiv \langle \sigma \rangle ^{2}/\langle \sigma ^{2}\rangle $
and $\beta $ is adjusted as to reproduce the exact third virial coefficient
for the mixture $B_{3}$, namely 
\begin{equation}
\beta =\frac{B_{3}}{(\pi /4)^{2}\langle \sigma ^{2}\rangle ^{2}(1+\xi )}-%
\frac{b_{3}}{2}.  \label{2}
\end{equation}
Here, $b_{3}=4(4/3-\sqrt{3}/\pi )$ is the reduced third virial coefficient
for the single component system while $B_{3}$ is given by \cite{BXHB88} 
\begin{equation}
B_{3}=\frac{\pi }{3}\left( a_{11}x_{1}^{3}\sigma
_{1}^{4}+3a_{12}x_{1}^{2}x_{2}\sigma _{12}^{4}+3a_{21}x_{1}x_{2}^{2}\sigma
_{12}^{4}+a_{22}x_{2}^{3}\sigma _{2}^{4}\right) ,  \label{3}
\end{equation}
where 
\begin{equation}
a_{ij}=\pi +2\left( \sigma _{i}^{2}/\sigma _{ij}^{2}-1\right) \cos
^{-1}(\sigma _{i}/2\sigma _{ij})-\sqrt{4\sigma _{ij}^{2}/\sigma _{i}^{2}-1}%
(1+\sigma _{i}^{2}/2\sigma _{ij}^{2})\sigma _{i}^{2}/2\sigma _{ij}^{2}
\label{4}
\end{equation}
and $\sigma _{ij}=(\sigma _{i}+\sigma _{j})/{2}$.

The EOS for the mixture, consistent with a given EOS for a single component
system that we introduced recently reads \cite{SYH99} 
\begin{equation}
Z_{\text{m}}^{\text{SYH}}(\eta )=\left( 1-\xi \right) \frac{1}{1-\eta }+\xi
Z_{\text{s}}(\eta ).  \label{5}
\end{equation}
We stress the fact that Eq. (\ref{5}) is simpler than Eq.\ (\ref{1}) [which
must be complemented with Eqs.\ (\ref{2})--(\ref{4})]. In addition, the
structure of Eq.\ (\ref{5}) is valid for any number of components, while the
third virial coefficient is known exactly only for binary mixtures.

In Table \ref{table1}, we show the results of Eqs.\ (\ref{1}) and (\ref{5})
when Woodcock's EOS \cite{W76} for the single component system [{cf. } Eq.\
(15) in Ref.\ \cite{BS01}] is used as input in both, as well as the
available MC data. Although not shown, we have also performed the comparison
using other EOS for the single component system, namely the one by Henderson
\cite{H77}, ours \cite{SHY95}, and the very accurate Levin(6) approximant of
Erpenbeck and Luban\cite{EL85}. In all instances, it is fair to say that
both recipes are of comparable accuracy with respect to the Monte Carlo
results, their difference being generally smaller than the error bars of the
simulation data. Nevertheless, in the particular case of using Woodcock's
EOS for $Z_{\text{s}}(\eta )$, as seen in Table \ref{table1}, $Z_{\text{m}}^{%
\text{BS}}(\eta )$ performs slightly better than $Z_{\text{m}}^{\text{SYH}%
}(\eta )$. This may be fortuitous since it is known that the Levin(6)
approximant gives the most accurate approximation to $Z_{\text{s}}(\eta )$ 
\cite{SHY95,EL85} and when using it in Eqs.\ (\ref{1}) and (\ref{5}) the
apparent (slight) superiority of $Z_{\text{m}}^{\text{BS}}(\eta )$ is no
longer there. For instance, the theoretical values of Table \ref{table1}
corresponding to the packing fraction $\eta =0.6$ are increased by about
0.03 when the Levin(6) approximant rather than Woodcock's EOS is used as
input, so that in this case the accuracy of $Z_{\text{m}}^{\text{SYH}}(\eta )
$ is slightly better than that of $Z_{\text{m}}^{\text{BS}}(\eta )$. 
\begin{table}[tbp]
\caption{Compressibility factor for different binary mixtures of hard disks
as obtained from Monte Carlo simulations, from Eq.\ ({\ref{1}}), and from
Eq.\ ({\ref{5}}). In the two latter, Woodcock's equation of state for the
single component system is used.}
\label{table1}
\begin{tabular}{ccccccccccc}
&  & \multicolumn{3}{c}{$x_1=0.25$} & \multicolumn{3}{c}{$x_1=0.50$} & 
\multicolumn{3}{c}{$x_1=0.75$} \\ \cline{3-5}\cline{6-8}\cline{9-11}
$\sigma_2/\sigma_1$ & $\eta$ & MC\tablenote{Ref.\ \protect\cite{BS01}} & 
Eq.\ (\ref{1}) & Eq.\ (\ref{5}) & MC$^{\text{a}}$ & Eq.\ (\ref{1}) & Eq.\ (%
\ref{5}) & MC$^{\text{a}}$ & Eq.\ (\ref{1}) & Eq.\ (\ref{5}) \\ 
\tableline $2/3$ & 0.20 & 1.559(6) & 1.559 & 1.559 & 1.561(5) & 1.558 & 1.558
& 1.565(4) & 1.563 & 1.563 \\ 
& 0.30 & 2.036(8) & 2.040 & 2.040 & 2.043(8) & 2.039 & 2.039 & 2.051(8) & 
2.048 & 2.048 \\ 
& 0.40 & 2.79(1) & 2.79 & 2.79 & 2.79(1) & 2.78 & 2.78 & 2.80(1) & 2.80 & 
2.80 \\ 
& 0.45 & 3.31(1) & 3.32 & 3.32 & 3.31(2) & 3.32 & 3.32 & 3.33(1) & 3.34 & 
3.34 \\ 
& 0.50 & 4.02(2) & 4.02 & 4.02 & 4.02(1) & 4.02 & 4.02 & 4.04(2) & 4.05 & 
4.05 \\ 
& 0.55 & 4.98(2) & 4.97 & 4.97 & 4.98(2) & 4.96 & 4.96 & 5.03(2) & 5.00 & 
5.00 \\ 
& 0.60 & 6.31(2) & 6.29 & 6.28 & 6.30(1) & 6.28 & 6.27 & 6.36(3) & 6.33 & 
6.33 \\ 
&  &  &  &  &  &  &  &  &  &  \\ 
$1/2$ & 0.20 & 1.534(6) & 1.536 & 1.536 & 1.540(6) & 1.538 & 1.538 & 1.556(7)
& 1.552 & 1.552 \\ 
& 0.30 & 1.998(7) & 1.995 & 1.995 & 2.008(7) & 2.000 & 2.000 & 2.039(8) & 
2.027 & 2.026 \\ 
& 0.40 & 2.72(1) & 2.70 & 2.70 & 2.71(1) & 2.71 & 2.71 & 2.77(1) & 2.76 & 
2.76 \\ 
& 0.45 & 3.20(2) & 3.21 & 3.21 & 3.22(2) & 3.22 & 3.22 & 3.29(2) & 3.29 & 
3.29 \\ 
& 0.50 & 3.88(1) & 3.88 & 3.87 & 3.90(2) & 3.89 & 3.89 & 3.98(2) & 3.98 & 
3.98 \\ 
& 0.55 & 4.79(2) & 4.77 & 4.76 & 4.81(2) & 4.80 & 4.78 & 4.93(2) & 4.91 & 
4.90 \\ 
& 0.60 & 6.03(3) & 6.02 & 6.00 & 6.04(2) & 6.05 & 6.03 & 6.22(3) & 6.21 & 
6.20 \\ 
&  &  &  &  &  &  &  &  &  &  \\ 
$1/3$ & 0.20 & 1.491(6) & 1.490 & 1.490 & 1.510(8) & 1.506 & 1.506 & 1.538(9)
& 1.536 & 1.536 \\ 
& 0.30 & 1.907(8) & 1.905 & 1.904 & 1.940(8) & 1.937 & 1.936 & 2.004(9) & 
1.996 & 1.995 \\ 
& 0.40 & 2.55(1) & 2.54 & 2.54 & 2.59(1) & 2.60 & 2.60 & 2.71(1) & 2.71 & 
2.70 \\ 
& 0.45 & 2.99(2) & 3.00 & 2.99 & 3.07(1) & 3.07 & 3.06 & 3.20(2) & 3.21 & 
3.21 \\ 
& 0.50 & 3.60(2) & 3.59 & 3.57 & 3.69(2) & 3.69 & 3.68 & 3.89(2) & 3.88 & 
3.87 \\ 
& 0.55 & 4.39(3) & 4.39 & 4.36 & 4.52(2) & 4.52 & 4.50 & 4.79(2) & 4.78 & 
4.76 \\ 
& 0.60 &  & 5.49 & 5.44 & 5.66(6) & 5.68 & 5.64 & 6.06(1) & 6.03 & 6.00
\end{tabular}
\end{table}

\begin{figure}[tbp]
\centerline{\parbox{\textwidth}{\epsfxsize=.7\hsize\epsfbox{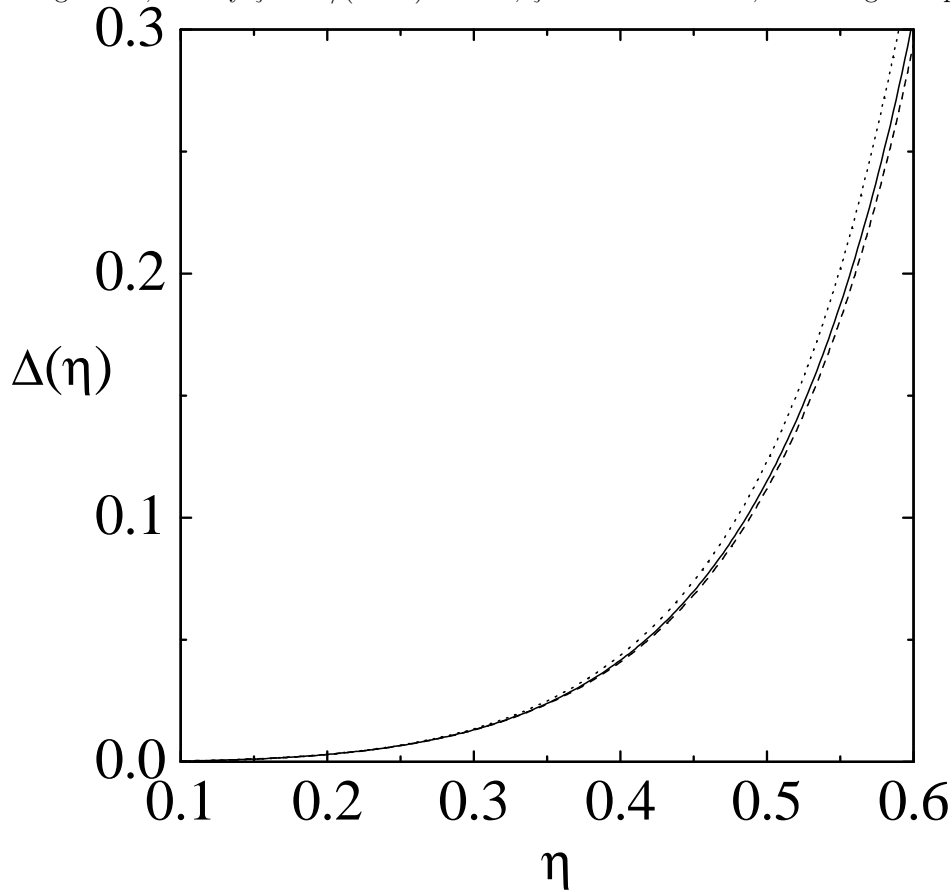}}}
\caption{Plot of $\Delta (\protect\eta )$ by assuming Henderson's EOS
(dotted line), Woodcock's EOS (dashed line), and the Levin(6) approximant
(solid line) for the pure fluid. }
\label{fig1}
\end{figure}

Let us try to understand why both EOS give practically equivalent results.
First, it may be shown that $Z_{\text{m}}^{\text{SYH}}(\eta )$, while not
reproducing the exact third virial coefficient $B_{3}$, yields a very good
estimate of it \cite{SYH01}. If we replace that estimate into Eq.\ (\ref{2}%
), we get 
\begin{equation}
\beta \simeq \frac{1-\xi }{1+\xi }\left( 1-\frac{b_{3}}{2}\right) .
\label{6}
\end{equation}
By using this estimate in Barrio and Solana's EOS, we have 
\begin{equation}
Z_{\text{m}}^{\text{BS}}(\eta )-Z_{\text{m}}^{\text{SYH}}(\eta )\simeq
\left( 1-\xi \right) \Delta (\eta ),  \label{7}
\end{equation}
where 
\begin{equation}
\Delta (\eta )\equiv \frac{1}{2}\left[ 1+\left( 1-\frac{b_{3}}{2}\right)
\eta \right] \left[ Z_{\text{s}}(\eta )-1\right] -\frac{\eta }{1-\eta }.
\label{8}
\end{equation}
According to the approximation involved in Eq.\ (\ref{7}), the difference $Z_{%
\text{m}}^{\text{BS}}(\eta )-Z_{\text{m}}^{\text{SYH}}(\eta )$ is small if
the asymmetry of the mixture is small ($\xi \lesssim 1$) and/or $\Delta
(\eta )$ is small. The function $\Delta (\eta )$ is plotted in Fig.\ \ref
{fig1} for the cases where $Z_{\text{s}}(\eta )$ is given by Henderson's
EOS, by Woodcock's EOS, and by the Levin(6) approximant. In all instances it
is practically zero up to $\eta \approx 0.2$ but then it grows rapidly. The
most disparate mixture considered in Barrio and Solana's simulations
corresponds to $x_{1}=0.25$, $\alpha \equiv \sigma _{2}/\sigma _{1}=1/3$,
which yields $\xi =0.75$. This explains the fact that $Z_{\text{m}}^{\text{BS%
}}(\eta )-Z_{\text{m}}^{\text{SYH}}(\eta )\lesssim 0.05$ in the simulated
cases. It should be noted however that the right-hand side of 
Eq.\ (\ref{7}) tends to overestimate
the actual difference $Z_{\text{m}}^{\text{BS}}(\eta )-Z_{\text{m}}^{\text{%
SYH}}(\eta )$, so that its main purpose is to illustrate the
fact that both EOS yield practically equivalent results for not very
asymmetric mixtures. On the other hand, more important differences can be
expected for disparate mixtures, especially in the case of large densities.
At a given density and a given diameter ratio $\alpha \leq 1$ the smallest
value of the parameter $\xi $ corresponds to a mole fraction $x_{1}=\alpha
/(1+\alpha )$ for the large disks, namely $\xi =4\alpha /(1+\alpha )^{2}$.
Thus, $\xi \ll 1$ if $\alpha \ll 1$ and, according to Eq.\ (\ref{7}), $Z_{%
\text{m}}^{\text{BS}}(\eta )-Z_{\text{m}}^{\text{SYH}}(\eta )\simeq \Delta
(\eta )$. 

Let us consider now the limit in which the small disks become point
particles ($\alpha\to 0$) and occupy a negligible fraction of the total
area. In that case, the compressibility factor of the mixture reduces to 
\cite{SYH99} 
\begin{equation}
Z_{\text{m}}(\eta)\to \frac{x_2}{1-\eta}+x_1 Z_{\text{s}}(\eta).  \label{9}
\end{equation}
The first term represents the (ideal gas) partial pressure due to the point
particles in the available area (i.e., the total area minus the area
occupied by the large disks), while the second term represents the partial
pressure associated with the large disks. In the limit $\alpha\to 0$ with $%
x_1$ finite (or, more generally, $x_1\gg\alpha$), we have $\xi\to x_1$ and $%
\beta\to (1-b_3/2)x_2/(1+x_1)$ [note in that limit the approximation (\ref{6}%
) is correct], so that 
\begin{equation}
Z_{\text{m}}^{\text{SYH}}(\eta)\to \frac{x_2}{1-\eta}+x_1 Z_{\text{s}}(\eta),
\label{10}
\end{equation}
\begin{equation}
Z_{\text{m}}^{\text{BS}}(\eta )\to 1+\frac{1}{2}\left[1+x_1+x_2\left(1-\frac{%
b_3}{2}\right)\eta\right] \left[ Z_{\text{s}}(\eta )-1\right] .  \label{11}
\end{equation}
Therefore, while Eq.\ (\ref{5}) is consistent with the exact property (\ref
{9}), Eq.\ (\ref{1}) violates it. In fact, the right-hand side of (\ref{7}),
with $\xi=x_1$, gives the deviation of Barrio and Solana's EOS from the
exact compressibility factor in the special case $\alpha\to 0$.

The research of M.L.H. was supported in part by DGAPA-UNAM under Project
IN103100. S.B.Y. and A.S. acknowledge partial support from the Ministerio de
Ciencia y Tecnolog\'{\i}a  (Spain) through grant No.\ BFM2001-0718. They are
also grateful to the DGES (Spain) for sabbatical grants No.\ PR2000-0116 and
No.\ PR2000-0117, respectively.

\end{document}